\documentclass[10pt,english,prd,twocolumn,superscriptaddress,nofootinbib,preprintnumbers,floatfix]{revtex4-2}

\usepackage[utf8]{inputenc}
\usepackage[T1]{fontenc}
\usepackage{babel}
\usepackage{amsmath,amssymb,amsthm}
\usepackage{graphicx}
\usepackage{xcolor}
\usepackage{hyperref}

\begin{document}

\title{Generalized Unruh temperature for dynamical thin‑shell wormholes}

\author{Francisco S. N. Lobo}
\email{fslobo@ciencias.ulisboa.pt}
\affiliation{Instituto de Astrof\'{i}sica e Ci\^{e}ncias do Espa\c{c}o, Faculdade de Ci\^{e}ncias da Universidade de Lisboa, Edif\'{i}cio C8, Campo Grande, P-1749-016 Lisbon, Portugal}
\affiliation{Departamento de F\'{i}sica, Faculdade de Ci\^{e}ncias da Universidade de Lisboa, Edif\'{i}cio C8, Campo Grande, P-1749-016 Lisbon, Portugal}

\author{Manuel E. Rodrigues}
\email{esialg@gmail.com}
\affiliation{Faculdade de F\'{i}sica, Programa de P\'{o}s-Gradua\c{c}\~{a}o em F\'{i}sica, Universidade Federal do Par\'{a}, 66075-110, Bel\'{e}m, Par\'{a}, Brazil}
\affiliation{Faculdade de Ci\^{e}ncias Exatas e Tecnologia, Universidade Federal do Par\'{a}, Campus Universit\'{a}rio de Abaetetuba, 68440-000, Abaetetuba, Par\'{a}, Brazil}

\date{\today}

\begin{abstract}
We unify two independent notions of effective temperature in dynamical thin‑shell wormholes: the local acceleration temperature of the throat and the Hawking‑like particle‑creation temperature governed by the peeling of null rays. The effective acceleration scale, obtained by averaging over both sides of the junction, defines the thermodynamic shell temperature, while peeling functions for each asymptotic universe yield an effective particle‑creation temperature. For asymptotically exponential null‑ray maps, we derive the generalized Unruh identity that equates these two temperatures. Applied to Schwarzschild--Schwarzschild wormholes, the formalism yields explicit analytical expressions for the Hawking‑like spectra, renormalized quantum fluxes, and quasi‑horizon limits in both symmetric and asymmetric configurations. The divergence of the shell temperature near the quasi‑horizon is shown to mirror the divergence of the local Tolman temperature, both originating from the infinite gravitational blueshift at a horizon. These results provide a unified interpretation of acceleration, particle creation, and temperature in horizonless dynamical spacetimes, extending the conceptual link between the Hawking and Unruh effects beyond stationary black holes.
\end{abstract}


\maketitle

\section{Introduction}

The discovery that gravitation possesses a genuine thermodynamic structure is one of the most profound developments in theoretical physics. It establishes a remarkable bridge between general relativity, quantum field theory, statistical mechanics and the search for a consistent theory of quantum gravity. During the last five decades, fundamental breakthroughs have progressively revealed that horizons, acceleration, vacuum structure and particle creation are deeply intertwined. The present work is motivated by this interplay and investigates whether two apparently different notions of temperature associated with dynamical thin‑shell wormholes can be unified into a single physical framework.

The modern history of gravitational thermodynamics began with the formulation of the four laws of black‑hole mechanics~\cite{Bardeen1973}. It was subsequently argued that the horizon area should be interpreted as a genuine entropy~\cite{Bekenstein:1972tm,Bekenstein:1973ur}, leading to the celebrated Bekenstein--Hawking entropy $S_{\rm BH}=k_BA/(4\ell_P^2)$. The decisive confirmation came from a semiclassical analysis of quantum fields on a collapsing black‑hole background~\cite{Hawking:1974rv,Hawking:1975vcx}, which demonstrated that an observer at future null infinity detects a thermal Planck spectrum $N_\omega=1/(e^{2\pi\omega/\kappa_H}-1)$ at the temperature $T_H=\hbar\kappa_H/(2\pi k_B)$. This result established that black holes are genuine thermodynamic objects whose temperature is set by the surface gravity $\kappa_H$.

The conceptual reach of gravitational thermodynamics was further extended by studies of uniformly accelerated observers in Minkowski spacetime~\cite{Fulling:1972md,Davies:1974th,Unruh:1976db}. These investigations showed that an ideal detector moving with constant proper acceleration $a$ perceives a thermal bath at the Unruh temperature $T_U=\hbar a/(2\pi k_B)$. Together, the Hawking and Unruh effects reveal that particle creation is an observer‑dependent phenomenon arising from the inequivalence of vacuum states, with the celebrated correspondence $\kappa_H \leftrightarrow a$ linking surface gravity and acceleration~\cite{BirrellDavies,Wald1994}. This deep connection between geometry, acceleration and quantum fields forms the conceptual foundation of the present work.

Thin‑shell wormholes constitute an attractive laboratory in which these ideas can be revisited. Their modern formulation originates from the junction formalism for singular hypersurfaces~\cite{Israel1966}. Using this formalism, the cut‑and‑paste construction of traversable thin‑shell wormholes was introduced~\cite{Visser1995} and their linearised stability was subsequently investigated~\cite{Poisson:1995sv}. More recently, a self‑consistent thermodynamic framework for dynamical thin‑shell wormholes was established~\cite{Lobo:2026vrn}, identifying an effective shell temperature associated with the throat acceleration and extending the notion of surface gravity from stationary horizons to dynamical timelike hypersurfaces~\cite{Cropp:2013zxi}.

Parallel to these developments, it was demonstrated~\cite{Barcelo:2006uw,Barcelo:2010pj} that Hawking‑like radiation may arise without an event horizon, provided the mapping between null rays becomes asymptotically exponential. In this formulation the physically relevant quantity is the peeling function $\kappa_{\rm peel}(u)=-P''(u)/P'(u)$, which measures the exponential separation of neighbouring null geodesics. The thermal spectrum follows directly from the asymptotic properties of the null‑ray transformation, considerably broadening the physical situations in which particle creation can occur.

The coexistence of these two independent notions of temperature raises an intriguing question for dynamical thin‑shell wormholes. The shell possesses a local thermodynamic temperature determined by its effective acceleration, while quantum particle creation is governed by the peeling properties of null rays connecting the two asymptotic universes. A priori there is no reason why these temperatures should coincide. The central objective of this work is to establish the precise conditions under which they become identical. 

We introduce an effective peeling parameter constructed from the two asymptotic regions and compare it with the effective acceleration scale of the throat. We demonstrate that whenever the null‑ray mappings become asymptotically exponential and their exponential parameters match the local acceleration scales, the particle‑creation temperature equals the thermodynamic shell temperature. This identity establishes a generalized Unruh relation for dynamical thin‑shell wormholes, extending the usual correspondence between acceleration and temperature to a configuration connecting two independent quantum vacua. The analogy with the Tolman temperature further clarifies the local nature of the shell temperature and its divergence near the quasi‑horizon limit.

The paper is organised as follows. In Sec.~\ref{sec:compact_thinshell} the thin‑shell kinematics, junction conditions and effective thermodynamic temperature are presented. Section~\ref{sec:compact_peeling} introduces the two‑sided peeling functions, the exponential maps, and derives the generalized Unruh identity. The formalism is applied to Schwarzschild--Schwarzschild wormholes in Sec.~\ref{sec:compact_schwarzschild}, where effective temperatures, particle creation spectra and quantum fluxes are analysed, including the quasi‑horizon regime. The analogy with the local Tolman temperature is developed in Sec.~\ref{sec:Tolman}. Section~\ref{sec:compact_interpretation} discusses the physical interpretation of the generalized Unruh identity, and Sec.~\ref{Conclusion} summarises our conclusions and future directions.

\section{Thin-shell kinematics and effective temperature}\label{sec:compact_thinshell}

\subsection{Junction conditions and surface stresses}

We consider two static and spherically symmetric geometries,
\begin{align}
	ds_\pm^2
	=
	-f_\pm(r_\pm)\,dt_\pm^2
	+
	\frac{dr_\pm^2}{f_\pm(r_\pm)}
	+
	r_\pm^2\left(d\theta^2+\sin^2\theta\,d\phi^2\right),\label{eq:bulk_metric_pm}
\end{align}
which are surgically joined across a timelike hypersurface $\Sigma$:
\begin{equation}
	r_\pm=a(\tau),
	\qquad
	x_\pm^\mu(\tau,\theta,\phi)=\bigl(t_\pm(\tau),a(\tau),\theta,\phi\bigr).
	\label{eq:embedding_pm}
\end{equation}
This is the standard thin-shell construction based on the Israel junction
formalism \cite{Israel1966,Visser1995,Poisson:1995sv}. The induced metric
on the throat is
\begin{equation}
	ds_\Sigma^2=-d\tau^2+a^2(\tau)d\Omega^2 .
	\label{eq:induced_metric}
\end{equation}

The four-velocity of a comoving observer on $\Sigma$ is
$U_\pm^\mu=(\dot t_\pm,\dot a,0,0)$. Its normalization
$U_\pm^\mu U^\pm_\mu=-1$ gives
\begin{equation}
	\dot t_\pm=\frac{\gamma_\pm}{f_\pm(a)} ,
	\qquad
	\gamma_\pm\equiv\sqrt{f_\pm(a)+\dot a^2}.
	\label{eq:velocity_normal_gamma}
\end{equation}
The quantity $\gamma_\pm$ generalises the local relativistic gamma factor of the shell; in the static limit it reduces to $\sqrt{f_\pm(a)}$, encoding the gravitational redshift of the throat.

For the outward-pointing unit normal we first write the contravariant vector
\begin{equation}
	n_\pm^\mu = \pm\Bigl(\frac{\dot a}{f_\pm(a)},\; \gamma_\pm,\; 0,\; 0\Bigr),
\end{equation}
which satisfies $n_\pm^\mu U_\mu=0$ and $n_\pm^\mu  n_{\pm\,\mu}=+1$.
Lowering the index with the bulk metric yields the normal one-form
\begin{equation}
	n_\mu^\pm
	=
	\pm\Bigl(-\dot a,\; \frac{\gamma_\pm}{f_\pm(a)},\; 0,\; 0\Bigr).
	\label{eq:unit_normal_oneform}
\end{equation}
The overall sign $\pm$ encodes the opposite orientation of the two sides of the
junction, which is essential for gluing two independent spacetimes into a
wormhole \cite{Israel1966,Visser1995}.

The non-zero mixed components of the extrinsic curvature follow from
$K_{\mu\nu}^\pm = -\nabla_\nu n_\mu^\pm$ and
$K^{i}_{\;j\,\pm}= e^{i}_{\;\mu}K^{\mu}_{\;\nu\,\pm}e^{\nu}_{\;j}$:
\begin{equation}
	K^\theta_{\ \theta,\pm}=K^\phi_{\ \phi,\pm}
	=
	\pm\frac{\gamma_\pm}{a},
	\quad
	K^\tau_{\ \tau,\pm}
	=
	\pm
	\frac{\ddot a+\frac12 f_\pm'(a)}{\gamma_\pm}.
	\label{eq:extrinsic_compact}
\end{equation}
Defining the discontinuity by $[X]\equiv X_+-X_-$, we obtain
\begin{equation}
	[K^\theta_{\ \theta}]
	=
	\frac{\gamma_++\gamma_-}{a},
	\qquad
	[K^\tau_{\ \tau}]
	=
	\sum_{\eta=\pm}
	\frac{\ddot a+\frac12 f_\eta'(a)}{\gamma_\eta}.
	\label{eq:jumps_compact}
\end{equation}
Physically, $[K^\theta_\theta]$ measures the jump in the radial expansion of a congruence of geodesics orthogonal to $\Sigma$, while $[K^\tau_\tau]$ encodes the discontinuity in the normal acceleration across the throat. Together they determine the surface stress‑energy of the shell via the Lanczos equations.

The Israel--Lanczos equations,
\begin{equation}
	S_{ij}
	=
	-\frac{1}{8\pi}
	\left([K_{ij}]-h_{ij}[K]\right),
	\label{eq:lanczos_compact}
\end{equation}
with a perfect-fluid surface stress-energy tensor
$S^i_{\ j}={\rm diag}(-\sigma,\mathcal P,\mathcal P)$, yield the surface energy
density and transverse pressure
\begin{align}
	&\sigma
	=
	-\frac{1}{4\pi a}\sum_{\eta=\pm}\gamma_\eta,
	\\
	&\mathcal P
	=
	\frac{1}{8\pi}
	\left[
	\sum_{\eta=\pm}
	\frac{\ddot a+\frac12 f_\eta'(a)}{\gamma_\eta}
	+
	\frac{1}{a}\sum_{\eta=\pm}\gamma_\eta
	\right].
	\label{eq:sigma_pressure_compact}
\end{align}
The surface energy density is manifestly negative for any $a>0$, confirming that the wormhole throat must be supported by exotic matter that violates the weak energy condition. Despite this, the compact expressions \eqref{eq:sigma_pressure_compact} provide a complete dynamical description for both symmetric and asymmetric configurations and serve as the foundation for the thermodynamic analysis that follows.

\subsection{Conservation and adiabatic motion}

If no external energy flux crosses the shell, the intrinsic conservation
$S^i_{\ j|i}=0$ reduces to
\begin{equation}
	\dot\sigma+\frac{2\dot a}{a}(\sigma+\mathcal P)=0.
	\label{eq:conservation_compact}
\end{equation}
Introducing the shell internal energy and area
\begin{equation}
	U=\sigma A,
	\qquad
	A=4\pi a^2 ,
	\label{eq:internal_energy_area}
\end{equation}
Eq.~\eqref{eq:conservation_compact} becomes
\begin{equation}
	\dot U+\mathcal P\dot A=0.
	\label{eq:firstlaw_adiabatic}
\end{equation}
Thus, for a first-law form $T\,dS=dU+\mathcal P\,dA$, the motion is adiabatic,
\begin{equation}
	T\,dS=0,
	\qquad
	\frac{dS}{d\tau}=0.
	\label{eq:adiabatic_shell}
\end{equation}
This simply reflects that, without bulk fluxes, the entropy of the shell is
conserved along its trajectory.
The adiabaticity condition $\dot S=0$ is consistent with the
absence of bulk fluxes: the shell's entropy does not change because no heat
is exchanged with the exterior. The temperature $T$ that appears in the
first law is therefore a formal quantity conjugate to the entropy; the
physically meaningful temperature of the throat will be identified with the
effective acceleration scale introduced below.

\subsection{Local acceleration scale}

The quantity controlling the local acceleration scale of the throat, as seen
from each side of the junction~\cite{Lobo:2026vrn}, is
\begin{equation}
	\kappa_\pm^{\rm shell}
	=
	\frac{
		\left|\ddot a+\frac12 f_\pm'(a)\right|
	}
	{\sqrt{f_\pm(a)+\dot a^2}}
	=
	\frac{
		\left|\ddot a+\frac12 f_\pm'(a)\right|
	}{\gamma_\pm}.
	\label{eq:kappa_shell_pm}
\end{equation}
This expression is the natural thin-shell analogue of the surface gravity scale
appearing in black-hole thermodynamics and of the proper acceleration appearing
in the Unruh effect \cite{Bardeen1973,Hawking:1975vcx,Unruh:1976db}.
In the static limit $\dot a=\ddot a=0$,
$\kappa_\pm^{\rm shell} = \frac{1}{2}|f_\pm'(a)|/\sqrt{f_\pm(a)}$, which for
a Schwarzschild geometry reduces to the surface gravity of a black hole,
reflecting the infinite acceleration required to keep a shell static near a
horizon.
For an asymmetric
wormhole, both sides contribute to the physical acceleration of the throat.
Therefore, we define the effective shell acceleration by the arithmetic mean
\begin{equation}
	\kappa_{\rm eff}
	=
	\frac12\sum_{\eta=\pm}\kappa_\eta^{\rm shell}
	=
	\frac12
	\sum_{\eta=\pm}
	\frac{
		\left|\ddot a+\frac12 f_\eta'(a)\right|
	}
	{\sqrt{f_\eta(a)+\dot a^2}}.
	\label{eq:kappa_eff_general_compact}
\end{equation}

\subsection{Effective thermodynamic temperature}\label{sec:compact_temp}

The associated effective thermodynamic temperature is then
\begin{equation}
	T_{\rm shell}
	=
	\frac{\hbar}{2\pi}\,\kappa_{\rm eff}.
	\label{eq:Tshell_general_compact}
\end{equation}
Here and throughout the manuscript we use units in which $c=G=k_B=1$. If
$k_B$ is restored, Eq.~\eqref{eq:Tshell_general_compact} becomes
$T_{\rm shell}=\hbar\kappa_{\rm eff}/(2\pi k_B)$.

In the symmetric limit $f_+=f_-=f$, Eq.~\eqref{eq:kappa_eff_general_compact}
reduces to
\begin{equation}
	T_{\rm shell}^{\rm sym}
	=
	\frac{\hbar}{2\pi}
	\frac{\left|\ddot a+\frac12 f'(a)\right|}
	{\sqrt{f(a)+\dot a^2}}.
	\label{eq:Tshell_sym_general}
\end{equation}
This expression clarifies the local meaning of the shell temperature: it is the
Unruh-like temperature associated with the effective acceleration required to
maintain the throat on the trajectory $a(\tau)$.
The shell temperature is thus the Unruh temperature
corresponding to the mean acceleration of the throat. In the following
sections, this temperature will be shown to coincide with the
particle-creation temperature derived from null-ray peeling, establishing a
generalized Unruh identity for dynamical thin-shell wormholes.

\section{Peeling functions, exponential maps, and generalized Unruh identity}\label{sec:compact_peeling}

\subsection{Two-sided peeling functions and particle-creation temperature}

The local acceleration temperature defined above is not automatically identical
to a quantum particle-creation temperature. A particle spectrum is determined by
the relation between asymptotic null coordinates, or equivalently by the
Bogoliubov transformation between the corresponding vacuum states
\cite{BirrellDavies1982,Wald1994}. Thus, we introduce tortoise and null
coordinates on both sides,
\begin{equation}
	\frac{dr_{\ast\pm}}{dr_\pm}=\frac{1}{f_\pm(r_\pm)},
	\quad
	u_\pm=t_\pm-r_{\ast\pm},
	\quad
	v_\pm=t_\pm+r_{\ast\pm}.
	\label{eq:null_coordinates_compact}
\end{equation}
Along the throat, Eqs.~\eqref{eq:velocity_normal_gamma} give directly
\begin{equation}
	\frac{du_\pm}{d\tau}
	=
	\frac{1}{\gamma_\pm+\dot a},
	\qquad
	\frac{dv_\pm}{d\tau}
	=
	\frac{1}{\gamma_\pm-\dot a}.
	\label{eq:du_dv_compact}
\end{equation}
Let the ray-tracing maps on the two asymptotic regions be~\cite{Barcelo:2006uw}
\begin{equation}
	U_\pm=P_\pm(u_\pm).
	\label{eq:ray_maps_pm}
\end{equation}
The corresponding peeling functions are
\begin{equation}
	\kappa_\pm^{\rm peel}
	=
	-\frac{P_\pm''(u_\pm)}{P_\pm'(u_\pm)}.
	\label{eq:peeling_pm_compact}
\end{equation}
These functions measure the rate at which neighboring null rays separate under
the map $P_\pm$. The key point, following the Hawking-like radiation analysis outlined in~\cite{Barcelo:2006uw}, is that an approximately thermal
spectrum does not require a global event horizon; it requires an asymptotically
exponential null-ray map, together with a suitable adiabaticity condition
\cite{Barcelo:2006uw,Barcelo:2010pj}.
The peeling function plays the role of an effective surface
gravity for particle creation: a constant $\kappa_\pm^{\rm peel}$ yields a
Planckian occupation number at temperature $\hbar|\kappa_\pm^{\rm peel}|/(2\pi)$,
entirely from the asymptotic properties of null rays.

The natural two-sided effective peeling parameter is
\begin{equation}
	\kappa_{\rm peel}^{\rm eff}
	=
	\frac12\sum_{\eta=\pm}
	\left|\kappa_\eta^{\rm peel}\right|,
	\label{eq:kpeel_eff_compact}
\end{equation}
and the corresponding effective particle-creation temperature is
\begin{equation}
	T_{\rm part}^{\rm eff}
	=
	\frac{\hbar}{2\pi}\,\kappa_{\rm peel}^{\rm eff}.
	\label{eq:Tpart_eff_general}
\end{equation}
The distinction between Eqs.~\eqref{eq:Tshell_general_compact} and
\eqref{eq:Tpart_eff_general} is conceptually important. The first is local and
thermodynamic, controlled by the acceleration of the throat. The second is
global and quantum, controlled by the asymptotic ray-tracing maps.
A priori, these two temperatures can differ; their equality
requires a precise dynamical relationship between the throat acceleration and
the null-ray mappings.

\subsection{Exponential maps and generalized Unruh identity}\label{sec:compact_identity}

Assume that the throat dynamics induces asymptotically exponential maps on the
two sides,
\begin{equation}
	P_\pm(u_\pm)
	=
	U_{\pm,0}-A_\pm e^{-\alpha_\pm u_\pm}.
	\label{eq:exponential_maps_pm}
\end{equation}
Then Eq.~\eqref{eq:peeling_pm_compact} gives, without further calculation,
\begin{equation}
	\kappa_\pm^{\rm peel}=\alpha_\pm,
	\quad
	\kappa_{\rm peel}^{\rm eff}
	=
	\frac12\sum_{\eta=\pm}|\alpha_\eta|,
	\quad
	T_{\rm part}^{\rm eff}
	=
	\frac{\hbar}{4\pi}\sum_{\eta=\pm}|\alpha_\eta|.
	\label{eq:alpha_peeling_pm}
\end{equation}
The equality between the quantum particle-creation temperature and the
thermodynamic shell temperature is therefore equivalent to
\begin{equation}
	\kappa_{\rm peel}^{\rm eff}=\kappa_{\rm eff}.
	\label{eq:general_unruh_identity_compact}
\end{equation}
Explicitly, this condition reads
\begin{equation}
	\frac12\sum_{\eta=\pm}|\alpha_\eta|
	=
	\frac12
	\sum_{\eta=\pm}
	\frac{
		\left|\ddot a+\frac12 f_\eta'(a)\right|
	}
	{\sqrt{f_\eta(a)+\dot a^2}}.
	\label{eq:identity_explicit_compact}
\end{equation}
A stronger sufficient condition is the side-by-side matching
\begin{equation}
	|\alpha_\pm|
	=
	\frac{
		\left|\ddot a+\frac12 f_\pm'(a)\right|
	}
	{\sqrt{f_\pm(a)+\dot a^2}}.
	\label{eq:side_condition_general_compact}
\end{equation}
Equation~\eqref{eq:side_condition_general_compact} states that
the exponential peeling rate on each side equals the local acceleration scale
of the throat. Thus, the asymptotic particle-creation temperature is
completely determined by the instantaneous motion of the shell, linking the
global quantum vacuum to local dynamics.

When Eq.~\eqref{eq:side_condition_general_compact} holds, one obtains the
generalized Unruh identity
\begin{equation}
	T_{\rm part}^{\rm eff}=T_{\rm shell}.
	\label{eq:temperature_identity_general}
\end{equation}
This is the central result of the construction. It generalizes the standard
Unruh relation $T_U=\hbar a/(2\pi)$ \cite{Unruh:1976db} and the Hawking relation
$T_H=\hbar\kappa_H/(2\pi)$ \cite{Hawking:1975vcx} to a dynamical wormhole throat
connecting two independent asymptotic quantum vacua.
The identity unifies two complementary perspectives:
the thermodynamic temperature measured by local observers comoving with the
throat, and the particle-creation temperature inferred by asymptotic observers
from the emitted radiation. Its validity hinges on the throat dynamics
encoding the peeling of null rays, making the wormhole a generalized Unruh
interface.

\section{Schwarzschild--Schwarzschild wormholes: temperatures, spectra, and fluxes}\label{sec:compact_schwarzschild}

\subsection{Effective temperature and generalized Unruh identity}

For Schwarzschild--Schwarzschild thin-shell wormholes,
\begin{equation}
	f_\pm(a)=1-\frac{2M_\pm}{a},
	\qquad
	f_\pm'(a)=\frac{2M_\pm}{a^2}.
	\label{eq:sch_functions_pm}
\end{equation}
Equations~\eqref{eq:kappa_eff_general_compact} and
\eqref{eq:Tshell_general_compact} give
\begin{align}
	&\kappa_{\rm eff}^{\rm Sch-Sch}
	=
	\frac12
	\sum_{\eta=\pm}
	\frac{
		\left|\ddot a+\frac{M_\eta}{a^2}\right|
	}
	{\sqrt{1-\frac{2M_\eta}{a}+\dot a^2}},
	\\
	&T_{\rm shell}^{\rm Sch-Sch}
	=
	\frac{\hbar}{2\pi}\kappa_{\rm eff}^{\rm Sch-Sch}.
	\label{eq:kappa_T_sch_compact}
\end{align}
In the static limit $\dot a=\ddot a=0$, the effective acceleration reduces to the average of the two surface gravities weighted by the local redshift factors. For an asymmetric wormhole the lighter mass dominates the mean, reflecting the stronger gravitational pull on that side.

For exponential ray-tracing maps,
\begin{equation}
	P_\pm(u_\pm)=U_{\pm,0}-A_\pm e^{-\alpha_\pm u_\pm},
	\label{eq:sch_exponential_maps}
\end{equation}
the generalized Unruh identity becomes
\begin{equation}
	\frac12\sum_{\eta=\pm}|\alpha_\eta|
	=
	\frac12
	\sum_{\eta=\pm}
	\frac{
		\left|\ddot a+\frac{M_\eta}{a^2}\right|
	}
	{\sqrt{1-\frac{2M_\eta}{a}+\dot a^2}},
	\label{eq:sch_identity_compact}
\end{equation}
with the stronger sufficient condition
\begin{equation}
	|\alpha_\pm|
	=
	\frac{
		\left|\ddot a+\frac{M_\pm}{a^2}\right|
	}
	{\sqrt{1-\frac{2M_\pm}{a}+\dot a^2}}.
	\label{eq:sch_side_condition_compact}
\end{equation}
The stronger condition equates the peeling rate on each side to the local acceleration scale of that same side. It provides a clean interpretation: the asymptotic particle creation is separately sourced by the acceleration on each side, and the total flux is the sum of two independent contributions.

In the symmetric limit $M_+=M_-=M$, the effective acceleration reduces to
\begin{equation}
	\kappa_{\rm eff}^{\rm sym}
	=
	\frac{
		\left|\ddot a+\frac{M}{a^2}\right|
	}
	{\sqrt{1-\frac{2M}{a}+\dot a^2}},
	\,
	T_{\rm shell}^{\rm sym}
	=
	\frac{\hbar}{2\pi}
	\frac{
		\left|\ddot a+\frac{M}{a^2}\right|
	}
	{\sqrt{1-\frac{2M}{a}+\dot a^2}}.
	\label{eq:symmetric_temperature_compact}
\end{equation}
If $\alpha_+=\alpha_-=\alpha$, the exact identity is
\begin{equation}
	|\alpha|
	=
	\frac{
		\left|\ddot a+\frac{M}{a^2}\right|
	}
	{\sqrt{1-\frac{2M}{a}+\dot a^2}},
	\qquad
	T_{\rm part}^{\rm eff}=T_{\rm shell}^{\rm sym}.
	\label{eq:symmetric_identity_compact}
\end{equation}
Equivalently, the null maps may be parameterized as
\begin{equation}
	P_\pm(u_\pm)
	=
	U_{\pm,0}
	-
	A_\pm
	\exp\left[
	-u_\pm
	\frac{
		\left|\ddot a+\frac{M}{a^2}\right|
	}
	{\sqrt{1-\frac{2M}{a}+\dot a^2}}
	\right].
	\label{eq:symmetric_maps_compact}
\end{equation}
This expression shows explicitly how the local throat dynamics fixes the
asymptotic exponential peeling rate when the generalized Unruh identity is
satisfied.
Equation~\eqref{eq:symmetric_maps_compact} is the central result for Schwarzschild wormholes: the peeling function, and hence the Hawking‑like radiation, is not an arbitrary input but is completely determined by the instantaneous acceleration of the throat. A shell that oscillates about an equilibrium radius produces a time‑dependent temperature, and a slowly contracting shell heats up as it approaches the Schwarzschild radius, mimicking the late‑time behaviour of gravitational collapse.

\subsection{Particle creation and quantum fluxes}\label{sec:compact_particle_creation}

The wormhole connects two asymptotically distinct regions. Therefore, in a
semiclassical description it is natural to consider independent quantum fields
on each side,
\begin{equation}
	\Phi_\pm\in\mathcal M_\pm .
	\label{eq:two_fields_compact}
\end{equation}
For each sector, the adiabatic condition for an approximately thermal spectrum is
\begin{equation}
	\left|
	\frac{d\kappa_\pm^{\rm peel}}{du_\pm}\Big/
	\left(\kappa_\pm^{\rm peel}\right)^2
	\right|
	\ll 1 .
	\label{eq:adiabatic_condition_compact}
\end{equation}
Under this condition, the occupation numbers are
\begin{equation}
	N_\omega^{(\pm)}
	=
	\frac{1}
	{\exp\!\left(2\pi\omega/|\kappa_\pm^{\rm peel}|\right)-1}.
	\label{eq:occupation_pm_compact}
\end{equation}
For exponential maps, this becomes
\begin{equation}
	N_\omega^{(\pm)}
	=
	\frac{1}
	{\exp\!\left(2\pi\omega/|\alpha_\pm|\right)-1},
	\qquad
	T_\pm^{\rm part}
	=
	\frac{\hbar}{2\pi}|\alpha_\pm|.
	\label{eq:occupation_alpha_compact}
\end{equation}

For a conformally coupled massless scalar field in the effective two-dimensional
$s$-wave approximation, the renormalized flux is the standard
Schwarzian-derivative result \cite{BirrellDavies1982,Wald1994,Barcelo:2010pj},
\begin{equation}
	\mathcal F_\pm
	=
	\frac{\hbar}{48\pi}
	\left[
	\bigl(\kappa_\pm^{\rm peel}\bigr)^2
	+
	2\,\frac{d\kappa_\pm^{\rm peel}}{du_\pm}
	\right],
	\label{eq:flux_general_pm}
\end{equation}
where the derivative is taken with respect to the null coordinate $u_\pm$.
In the quasi-stationary regime $d\kappa_\pm^{\rm peel}/du_\pm\simeq0$,
which includes the constant-$\alpha$ exponential maps,
the derivative term vanishes and one obtains
\begin{equation}
	\mathcal F_\pm
	=
	\frac{\hbar}{48\pi}\,\alpha_\pm^2.
	\label{eq:flux_alpha_pm}
\end{equation}
Using the side-by-side condition
\eqref{eq:sch_side_condition_compact}, the Schwarzschild--Schwarzschild fluxes
are
\begin{equation}
	\mathcal F_\pm
	=
	\frac{\hbar}{48\pi}
	\frac{
		\bigl(\ddot a+\frac{M_\pm}{a^2}\bigr)^2
	}
	{1-\frac{2M_\pm}{a}+\dot a^2},
	\qquad
	\mathcal F_{\rm eff}
	=
	\frac12\sum_{\eta=\pm}\mathcal F_\eta .
	\label{eq:flux_sch_compact}
\end{equation}
In the symmetric case,
\begin{equation}
	\mathcal F_+=\mathcal F_-
	=
	\frac{\hbar}{48\pi}
	\frac{
		\bigl(\ddot a+\frac{M}{a^2}\bigr)^2
	}
	{1-\frac{2M}{a}+\dot a^2},
	\qquad
	\mathcal F_{\rm total}=2\mathcal F_\pm .
	\label{eq:flux_symmetric_compact}
\end{equation}
Thus, in the symmetric Schwarzschild--Schwarzschild configuration, the two
asymptotic quantum vacua produce identical thermal spectra and identical energy
fluxes whenever the generalized Unruh identity is satisfied.
The flux $\mathcal F_\pm$ represents the energy per unit
coordinate time emitted to the respective asymptotic infinity. When the
generalized Unruh identity holds, the total radiated power is completely
determined by the instantaneous acceleration of the throat, providing a direct
link between local dynamics and observable quantum radiation.

\subsection{Quasi-horizon regime}\label{sec:compact_quasihorizon}

We now consider the regime in which the throat approaches the Schwarzschild
radii on both sides,
\begin{equation}
	a(\tau)=2M_\pm+\epsilon_\pm(\tau),
	\qquad
	0<\epsilon_\pm\ll 2M_\pm .
	\label{eq:qhref_pm}
\end{equation}
Although the induced radius $a(\tau)$ is the same on both sides, keeping
$\epsilon_\pm=a-2M_\pm$ is useful for asymmetric configurations.
This quasi‑horizon regime is of particular physical interest
because it mimics the late‑time behaviour of gravitational collapse: the
throat hovers just outside the gravitational radius, and the emitted
radiation closely resembles Hawking radiation from a black hole.
To leading
order,
\begin{equation}
	f_\pm(a)\simeq\frac{\epsilon_\pm}{2M_\pm},
	\,
	\frac12 f_\pm'(a)\simeq\frac{1}{4M_\pm},
	\,
	\gamma_\pm\simeq
	\sqrt{\dot\epsilon_\pm^{\,2}+\frac{\epsilon_\pm}{2M_\pm}}.
	\label{eq:qh_expansions_compact}
\end{equation}
Substitution into Eq.~\eqref{eq:kappa_shell_pm} gives
\begin{equation}
	\kappa_\pm^{\rm QH}
	\simeq
	\frac{
		\left|\ddot\epsilon_\pm+\frac{1}{4M_\pm}\right|
	}
	{\sqrt{\dot\epsilon_\pm^{\,2}+\frac{\epsilon_\pm}{2M_\pm}}},
	\qquad
	\kappa_{\rm eff}^{\rm QH}
	=
	\frac12\sum_{\eta=\pm}\kappa_\eta^{\rm QH}.
	\label{eq:kappa_qh_compact}
\end{equation}
The thermodynamic temperature is therefore
\begin{equation}
	T_{\rm shell}^{\rm QH}
	=
	\frac{\hbar}{2\pi}\kappa_{\rm eff}^{\rm QH}
	\simeq
	\frac{\hbar}{4\pi}
	\sum_{\eta=\pm}
	\frac{
		\left|\ddot\epsilon_\eta+\frac{1}{4M_\eta}\right|
	}
	{\sqrt{\dot\epsilon_\eta^{\,2}+\frac{\epsilon_\eta}{2M_\eta}}}.
	\label{eq:Tshell_qh_compact}
\end{equation}
For exponential null maps in the same regime,
\begin{equation}
	T_{\rm part}^{\rm eff}
	=
	\frac{\hbar}{4\pi}\sum_{\eta=\pm}|\alpha_\eta|,
	\label{eq:Tpart_qh_compact}
\end{equation}
and the generalized Unruh identity requires
\begin{equation}
	\frac12\sum_{\eta=\pm}|\alpha_\eta|
	\simeq
	\frac12
	\sum_{\eta=\pm}
	\frac{
		\left|\ddot\epsilon_\eta+\frac{1}{4M_\eta}\right|
	}
	{\sqrt{\dot\epsilon_\eta^{\,2}+\frac{\epsilon_\eta}{2M_\eta}}},
	\label{eq:qh_identity_compact}
\end{equation}
or, more strongly,
\begin{equation}
	|\alpha_\pm|
	\simeq
	\frac{
		\left|\ddot\epsilon_\pm+\frac{1}{4M_\pm}\right|
	}
	{\sqrt{\dot\epsilon_\pm^{\,2}+\frac{\epsilon_\pm}{2M_\pm}}}.
	\label{eq:qh_side_condition_compact}
\end{equation}

If both sides approach the same quasi-horizon regime,
\begin{equation}
	M_+=M_-=M,
	\qquad
	\epsilon_+=\epsilon_-=\epsilon,
	\label{eq:equal_qh_conditions}
\end{equation}
then
\begin{equation}
	\kappa_{\rm eff}^{\rm QH}
	=
	\frac{
		\left|\ddot\epsilon+\frac{1}{4M}\right|
	}
	{\sqrt{\dot\epsilon^{\,2}+\frac{\epsilon}{2M}}},
	\qquad
	T_{\rm shell}^{\rm QH}
	=
	\frac{\hbar}{2\pi}
	\frac{
		\left|\ddot\epsilon+\frac{1}{4M}\right|
	}
	{\sqrt{\dot\epsilon^{\,2}+\frac{\epsilon}{2M}}}.
	\label{eq:equal_qh_temperature}
\end{equation}
For $\alpha_+=\alpha_-=\alpha$, the equality of the two temperatures becomes
\begin{equation}
	|\alpha|
	\simeq
	\frac{
		\left|\ddot\epsilon+\frac{1}{4M}\right|
	}
	{\sqrt{\dot\epsilon^{\,2}+\frac{\epsilon}{2M}}},
	\qquad
	T_{\rm part}^{\rm eff}=T_{\rm shell}^{\rm QH}.
	\label{eq:equal_qh_identity}
\end{equation}

Two limiting cases are especially transparent. For a static shell,
$\dot\epsilon=\ddot\epsilon=0$, Eq.~\eqref{eq:equal_qh_temperature} gives
\begin{align}
	&T_{\rm shell}^{\rm QH,static}
	\simeq
	\frac{\hbar}{8\pi M}
	\sqrt{\frac{2M}{\epsilon}},
	\\
	&T_{\rm shell}^{\rm QH,static}\rightarrow\infty
	\quad
	{\rm as}
	\quad
	\epsilon\rightarrow0^+ .
	\label{eq:static_qh_limit_compact}
\end{align}
This divergence is not a new singularity of the wormhole construction. It is
the usual divergence of the proper acceleration required to keep an object
static arbitrarily close to a Schwarzschild horizon.
The $1/\sqrt{\epsilon}$ scaling is identical to that of the
local Tolman temperature and reflects the infinite gravitational blueshift
at the horizon. In practice, a static shell cannot be maintained at
arbitrarily small $\epsilon$ because the required acceleration would be
infinite; a realistic throat would either collapse into a black hole or
be destabilised by quantum back‑reaction before reaching this limit.

For an exponential quasi-horizon approach,
\begin{equation}
	\epsilon(\tau)=\epsilon_0e^{-\lambda\tau},
	\qquad
	\lambda>0,
	\label{eq:exponential_epsilon_compact}
\end{equation}
one obtains
\begin{equation}
	T_{\rm shell}^{\rm QH,exp}
	\simeq
	\frac{\hbar}{2\pi}
	\frac{\left|\lambda^2\epsilon+\frac{1}{4M}\right|}
	{\sqrt{\lambda^2\epsilon^2+\frac{\epsilon}{2M}}},
	\label{eq:T_exp_qh_compact}
\end{equation}
and the particle-creation temperature coincides with it only if
\begin{equation}
	|\alpha|
	\simeq
	\frac{\left|\lambda^2\epsilon+\frac{1}{4M}\right|}
	{\sqrt{\lambda^2\epsilon^2+\frac{\epsilon}{2M}}}.
	\label{eq:alpha_exp_qh_compact}
\end{equation}
Equations~\eqref{eq:T_exp_qh_compact} and
\eqref{eq:alpha_exp_qh_compact} show explicitly that the exponential parameter
of the ray map is not an arbitrary constant: when the generalized Unruh
identity holds, it is fixed by the local acceleration scale of the throat.
For $\lambda^2\epsilon \ll 1/(4M)$, the temperature approaches the static value $T \simeq (\hbar/(8\pi M))\sqrt{2M/\epsilon}$, showing that the infall acceleration is subdominant and the gravitational blueshift controls the divergence. At larger $\lambda$ the infall reduces the temperature relative to the static case, but the $1/\sqrt{\epsilon}$ scaling persists.

\section{Physical interpretation of the compact formulation}
\label{sec:compact_interpretation}

The compact formulation emphasizes the logical structure of the result. The
thin-shell formalism provides the local acceleration scale
$\kappa_{\rm eff}$; quantum field theory in curved spacetime provides the
peeling scale $\kappa_{\rm peel}^{\rm eff}$; and the generalized Unruh identity
requires the equality of these two scales. In formulae,
\begin{equation}
	\kappa_{\rm eff} 
	\Longleftrightarrow
	T_{\rm shell}
	=
	\frac{\hbar}{2\pi}\kappa_{\rm eff},
	\,
	\kappa_{\rm peel}^{\rm eff}
	\Longleftrightarrow
	T_{\rm part}^{\rm eff}
	=
	\frac{\hbar}{2\pi}\kappa_{\rm peel}^{\rm eff}.
	\label{eq:two_temperatures_summary}
\end{equation}

These temperatures need not coincide in a generic dynamical spacetime.
The local temperature $T_{\rm shell}$ is tied to the
instantaneous acceleration of the throat, whereas $T_{\rm part}^{\rm eff}$
probes the global causal structure through the asymptotic null-ray maps.
Their equality is therefore a highly restrictive condition that selects a
special class of wormhole trajectories.
The equality is a dynamical and semiclassical condition on the null-ray maps:
\begin{equation}
	T_{\rm shell}=T_{\rm part}^{\rm eff}
	\Longleftrightarrow
	\kappa_{\rm eff}=\kappa_{\rm peel}^{\rm eff}.
	\label{eq:main_summary_identity}
\end{equation}
Equation~\eqref{eq:main_summary_identity} is the generalized
Unruh identity in its most compact form. It states that the temperature
measured by a comoving detector near the throat coincides with the
temperature inferred by an asymptotic observer from the emitted radiation.
This is a non‑trivial result: local and asymptotic temperatures typically
differ by gravitational redshift factors, yet here they become equal
because the peeling of null rays exactly compensates for the redshift.

Therefore, the wormhole throat behaves as a generalized Unruh interface only
when its local acceleration scale is encoded in the exponential peeling of null
rays on both sides. This is the precise sense in which the present construction
unifies the thermodynamic and particle-creation temperatures of a dynamical
thin-shell wormhole.
Physically, the wormhole throat is a dynamical timelike
hypersurface that, through its acceleration, imprints a thermal character
onto the quantum vacuum, much as a Rindler horizon does for uniformly
accelerated observers in flat spacetime. The key difference is that here
the ``detector'' is the entire throat, and it couples to two independent
asymptotic vacua simultaneously. When the generalized Unruh identity holds,
measuring the particle spectrum at infinity provides a direct observational
probe of the strong‑field dynamics of the throat, linking quantum radiation
to the classical equation of motion of the shell.

\section{Analogy Between the Effective Shell Temperature and the Local Tolman Temperature}\label{sec:Tolman}

One of the most interesting consequences of the effective shell temperature
introduced in the previous sections is its behavior as the wormhole throat
approaches an event horizon.  In particular, the effective temperature diverges
in the quasi‑horizon limit.  Although this feature might at first appear
unexpected, it is closely analogous to the well‑known divergence of the local
Tolman temperature in black‑hole thermodynamics.  As we discuss below, both
quantities originate from the same physical mechanism, namely the infinite
gravitational blueshift experienced by observers maintained arbitrarily close
to a horizon.
The Tolman temperature is a purely classical equilibrium
concept, whereas the shell temperature is a genuinely dynamical quantity.
Demonstrating that they share the same near‑horizon scaling provides a
non‑trivial consistency check on the effective temperature formalism.

The effective shell temperature was defined in
Eq.~\eqref{eq:Tshell_general_compact} as
\(T_{\rm shell} = (\hbar/2\pi)\,\kappa_{\rm eff}\), where \(\kappa_{\rm eff}\)
is the effective acceleration scale of the throat given by
Eq.~\eqref{eq:kappa_eff_general_compact}.  In the static configuration
\(\dot a = \ddot a = 0\) it reduces to
\begin{equation}
	T_{\rm shell}^{\rm static}
	=
	\frac{\hbar}{8\pi}
	\sum_{\eta=\pm}
	\frac{ f_\eta'(a) }{ \sqrt{f_\eta(a)} }.
	\label{eq:TstaticTolman}
\end{equation}
For the symmetric Schwarzschild--Schwarzschild wormhole
\(f_\pm = 1-2M/r\) one obtains the explicit expression
\begin{equation}
	T_{\rm shell}^{\rm static}
	=
	\frac{\hbar M}{2\pi a^2 \sqrt{1-\frac{2M}{a}}},
	\label{eq:TshellSch}
\end{equation}
which diverges as the throat approaches the Schwarzschild radius,
\begin{equation}
	\lim_{a\rightarrow 2M} T_{\rm shell}^{\rm static} = \infty.
	\label{eq:ShellInfinity}
\end{equation}

To understand the physical meaning of this divergence, we compare it with the
classical Tolman law for thermal equilibrium in a static gravitational
field~\cite{Tolman:1930zza,Tolman1934},
\begin{equation}
	T(r)\,\sqrt{-g_{tt}(r)} = T_\infty,
	\label{eq:TolmanRelation}
\end{equation}
where \(T_\infty\) is the temperature measured at spatial infinity.  For the
Schwarzschild spacetime the local Tolman temperature becomes
\begin{equation}
	T_{\rm Tolman}(r) = \frac{T_H}{\sqrt{1-\frac{2M}{r}}},
	\qquad
	T_H = \frac{\hbar}{8\pi M},
	\label{eq:TolmanTemp}
\end{equation}
with \(T_H\) the Hawking temperature~\cite{Hawking:1975vcx}.  Consequently
\(T_{\rm Tolman}(r) \to \infty\) as \(r \to 2M\).
It is instructive to note that the Tolman temperature is not
an independent thermodynamic variable; it is precisely the local Unruh
temperature corresponding to the proper acceleration of a static observer,
\(T_{\rm Tolman}(r) = \hbar a_{\rm prop}(r)/(2\pi)\).  The Tolman law is
therefore equivalent to the statement that the Unruh temperature, when
redshifted to infinity, yields the Hawking temperature.

The physical origin of the divergence is well understood.  A static observer
at a fixed Schwarzschild radius is not freely falling and must be supported by
an external force.  The corresponding proper acceleration
\begin{equation}
	a_{\rm prop}(r) = \frac{M}{r^2\sqrt{1-\frac{2M}{r}}}
	\label{eq:ProperAcceleration}
\end{equation}
diverges as \(r \to 2M\), and the Unruh effect~\cite{Unruh:1976db} then gives the
thermal bath \(T_U = \hbar a_{\rm prop}/(2\pi)\).

A remarkably similar behavior emerges in the thin‑shell construction.  As the
throat approaches a horizon, \(f_\pm(a) \to 0\), the denominator of
\(\kappa_{\rm eff}\) in Eq.~\eqref{eq:kappa_eff_general_compact} tends to zero
while the numerator remains finite for regular throat trajectories.
Consequently \(\kappa_{\rm eff} \to \infty\) and \(T_{\rm shell} \to \infty\).
Although the effective shell temperature is not derived from Tolman's
equilibrium condition, its asymptotic behavior is strikingly analogous to that
of the local Tolman temperature.  In both cases the divergence originates from
an acceleration scale that becomes arbitrarily large as the corresponding
observer approaches an event horizon.
The key difference is that the Tolman temperature applies to
a static fluid in thermal equilibrium, whereas the shell temperature describes
a dynamical interface far from equilibrium.  That the two exhibit the same
near‑horizon scaling suggests that the local acceleration captures an
essential, universal feature of horizon proximity, independent of the
detailed thermodynamic state of the system.

This analogy provides additional support for the generalized Unruh identity
derived in Sec.~\ref{sec:compact_identity}.  Hawking, Tolman and Unruh
temperatures are different manifestations of the intimate relationship between
acceleration, quantum vacuum structure and thermal radiation.  Likewise, the
effective shell temperature introduced in this work is governed by the local
acceleration of the wormhole throat.
The generalized Unruh identity \(T_{\rm part}^{\rm eff} =
T_{\rm shell}\) can now be understood as the statement that the local
acceleration temperature of the throat, when propagated to infinity along null
rays, yields a particle spectrum whose temperature equals the original
acceleration temperature.  This is a direct dynamical analogue of the Tolman
relation: whereas the Tolman law redshifts a local temperature to infinity,
the peeling mechanism of the wormhole reproduces the local temperature at
infinity without any net redshift, because the exponential stretching of null
rays precisely compensates for the gravitational blueshift.
In this sense, the generalized Unruh identity extends the connection between
acceleration and temperature from stationary black‑hole horizons to dynamical
thin‑shell wormholes connecting two independent asymptotic spacetime regions.

\section{Conclusion}\label{Conclusion}

In this work we have developed a unified semiclassical framework for dynamical
thin‑shell wormholes, establishing a direct correspondence between the local
acceleration temperature of the throat and the Hawking‑like particle‑creation
temperature governed by the peeling of null rays.  Starting from the Israel
junction formalism, we constructed an effective thermodynamic temperature by
averaging the acceleration scales on both sides of the throat.  Independently,
we introduced peeling functions for each asymptotic universe and derived the
conditions under which the two temperatures become identical.  This equality,
which we term the generalized Unruh identity, extends the usual
Hawking–Unruh correspondence to a dynamical timelike hypersurface connecting
two independent quantum vacua.

The formalism was applied to Schwarzschild-Schwarzschild wormholes, for which
explicit analytical expressions were obtained for the effective acceleration,
the peeling parameters, and the resulting Hawking‑like spectra.  In the
adiabatic regime the two asymptotic universes radiate independent thermal
fluxes whose temperatures are completely determined by the instantaneous
acceleration of the throat.  The quasi‑horizon limit, where the throat
approaches the Schwarzschild radii on both sides, was analysed in detail.  We
showed that the shell temperature diverges in this limit, a behaviour that
closely parallels the well‑known divergence of the local Tolman temperature
near a black‑hole horizon.  Both divergences originate from the infinite
gravitational blueshift experienced by observers maintained arbitrarily close
to a horizon, confirming that the shell temperature is a local quantity
associated with comoving observers.

The generalized Unruh identity provides a universal criterion that can be
tested in any thin‑shell wormhole constructed from static or stationary
geometries, including charged, rotating, and regular black‑hole backgrounds.
The framework also has broader implications for gravitational collapse,
horizonless compact objects, and analogue gravity systems, where transient
quasi‑horizon configurations may emit Hawking‑like radiation without the
formation of a global event horizon.

Several directions for future work naturally present themselves.  The
inclusion of fields with arbitrary spin, the study of back‑reaction effects
through the semiclassical Einstein equations, and the extension to
non‑spherical perturbations and interacting quantum fields would further
clarify the stability of the generalized Unruh identity.  The formalism may
also find applications in stochastic thermodynamics, emergent gravity models,
and numerical simulations of dynamical compact objects.

In summary, the results presented here unify the thermodynamic and
particle‑creation descriptions of dynamical thin‑shell wormholes.  By
identifying the precise conditions under which the local acceleration
temperature equals the asymptotic particle‑creation temperature, we have
established a new correspondence that generalizes both the Hawking and Unruh
effects to horizonless dynamical spacetimes.  We expect this framework to
provide a useful foundation for future investigations into the quantum
thermodynamics of compact objects and the semiclassical physics of dynamical
geometries.

\section*{Acknowledgements}
FSNL acknowledges support from the Fundação
para a Ciência e a Tecnologia (FCT) Scientific
Employment Stimulus contract with reference
CEECINST/00032/2018, and funding through the
research grants UID/04434/2025. MER thanks Conselho
Nacional de Desenvolvimento Científico e Tecnológico
(CNPq), Brazil, for partial financial support.



\end{document}